\documentclass[
 reprint,
 onecolumn,
 amsmath,amssymb,
 pra,
]{revtex4-2}

\newcommand{\ket}[1]{|#1\rangle}
\newcommand{\bra}[1]{\langle #1|}
\usepackage{graphicx}
\usepackage{placeins}

\usepackage{physics}
\usepackage[]{qcircuit}

\usepackage{amsthm}
\usepackage{amssymb}

\newtheorem{theorem}{Theorem}[section]
\newtheorem{corollary}{Corollary}[theorem]
\newtheorem{lemma}[theorem]{Lemma}

\usepackage[figurename=FIG., justification=RaggedRight]{caption}
\usepackage{subcaption} 
\usepackage[titletoc]{appendix}

\begin{document}

\author{D. V. Babukhin$^{1}$}

\affiliation{$^1$Dukhov Research Institute of Automatics (VNIIA), 127055 Moscow, Russia}

\title{Harrow-Hassidim-Lloyd algorithm without ancilla postselection}

\begin{abstract}
Harrow-Hassidim-Lloyd algorithm (HHL) allows for the exponentially faster solution of a system of linear equations.
However, this algorithm requires the postselection of an ancilla qubit to obtain the solution. This postselection makes the algorithm result probabilistic.
Here we show conditions when the HHL algorithm can work without postselection of ancilla qubit.
We derive expectation values for an observable $M$ on the HHL outcome state when ancilla qubit is measured in $\ket{0}$ and $\ket{1}$ and show condition for postselection-free HHL running.
We provide an explicit example of a practically-interesting input matrix and an observable, which satisfy postselection-free HHL condition.
Our work can improve the performance of the HHL-based algorithms.

\end{abstract}

\maketitle
\date{\today }

%
%

\section{Introduction}

The HHL algorithm \cite{Harrow2009} is a quantum algorithm that provides a solution for a linear system with an exponential speed-up. 
While having several caveats \cite{Aaronson2015}, this algorithm is an important subroutine in quantum computing.
Since its invention, numerous applications of this algorithm to practical problems have been demonstrated: solving linear systems is used for differential equations \cite{Cao2013, Berry2014}, calculating scattering cross-sections \cite{Clader2013}, and building quantum machine learning algorithms \cite{Rebentrost2014, Kerenidis2016, Wiebe2016}, or using this algorithm as a tool to attack cryptographic protocols \cite{Chen2017,Chen2018,Liu2021}. 
There are efforts to implement the HHL algorithm in a digital-analog approach to quantum computing \cite{Martin2022}, which is more resilient to quantum noise.
Progress in quantum computing devices in the last decades allowed conducting low-dimensional experiments, which investigate practical opportunities and caveats of the HHL algorithm \cite{Barz2014, Pan2014, Cai2013}. 

The HHL algorithm requires postselection of an ancilla qubit in a quantum state $\ket{1}$ to produce a solution. Ancilla measurement in the $\ket{1}$ state has a non-unity probability, which leads to discarding part of the algorithm runs on a quantum processor. Consequently, discarding results leads to an increase in quantum processor running time. An amplitude amplification algorithm \cite{Brassard2002} is usually used after the HHL circuit to increase the probability of measuring ancilla in $\ket{1}$.
This step requires $O(\kappa)$ repetitions of the amplitude amplification to make a success probability sufficiently high. Here $\kappa$ is a conditional number of the input matrix $A$, where $A$ represents a system of linear equations we want to solve. A running time of the HHL is $O(log(N)s^{2}\kappa^{2}/\epsilon)$\cite{Harrow2009}, where one $\kappa$ comes from the amplitude amplification step.
Although adding only a polynomial (in $\kappa$) complexity overhead, the amplitude amplification step increases algorithm running time and introduces gate errors into computation when the algorithm is running on a NISQ device \cite{Preskill2018}. Until fault-tolerant quantum computation is available, the postselection step decreases the efficiency of the HHL algorithm.

There were several works, which improve solution of linear systems on a quantum computer. In \cite{Ambainis2010}, a variable-time version of the amplitude amplification algorithm allows reducing dependence on conditional number from $\kappa^{2}$ to $\kappa \log^{3}{\kappa}$. 
In \cite{Childs2017}, a decomposition into a linear combination of unitary operations allows reducing dependence on precision from $1/\epsilon$ to $log(1/\epsilon)$. 
In \cite{Wossnig2018}, an algorithm for solving linear systems based on a quantum singular value estimation algorithm allows solving the problem for dense matrices. Finally, in recent paper \cite{Costa2022}, an algorithm for solving linear systems based on a discrete adiabatic theorem allows solving the linear problem with optimal query complexity of $O(\kappa \log(1/\epsilon))$.  
Still, there is surprising room for improvement in the original HHL algorithm.

In this paper, we demonstrate conditions for running the HHL algorithm without postselection of the ancilla qubit. This is possible for input matrices $A$ and for measurement matrices $M$, which satisfy a particular commutator identity. When this identity is satisfied, the algorithm produces quantum states for two ancilla measurement outcomes ($\ket{0}$ or $\ket{1}$), in which expectation values deviate from each other only by an easily-accessible constant. This connection of expectation values allows using both output states to obtain an expectation value of $M$ on the solution of the linear system. We provide an explicit example of an input matrix $A$, which satisfies the postselection-free condition and which is widely used in applications.
This reduction of postselection leads to the economy of $O(\kappa)$ operations of amplitude amplification, otherwise used to amplify the success probability of ancilla measurement. 

This paper is organized as follows. 
In Sec. II, we provide a formulation of the HHL protocol and derive the postselection-free condition. 
In Sec III, we demonstrate a practically-interesting example of input matrices and observables, which satisfy the postselection-free condition.
Finally, we conclude this paper and discuss perspectives in Sec. IV.
An appendix is devoted to a detailed workthrough of a toy example with a $2 \cross 2$ matrix, which demonstrates the postselection-free HHL.

%
%

\section{Derivation of the postselection-free condition}

\subsection{HHL algorithm}

Here we formulate the original HHL algorithm. The HHL algorithm \cite{Harrow2009} is designed to provide a solution for a system of linear equations in a form of a quantum state. Every system of linear equations can be represented with a coefficient matrix $A$, a vector of system solution $\Vec{x}$ and an income vector $\Vec{b}$. If a matrix $A$ is invertable, a unique solution for the linear system of equations exists. For an invertable matrix $A$, a vector $\Vec{x}$ of unknown variables and a vector $\Vec{b}$ of known values, a system of linear equations has a form
\begin{equation}
    A\Vec{x} = \Vec{b}
\end{equation}
with a solution of this system
\begin{equation}
    \label{classicalx}
    \Vec{x} = A^{-1}\Vec{b}.
\end{equation}
In quantum formalism, this solution has a form
\begin{equation}
    \label{quantumx}
    \ket{x} = A^{-1}\ket{b} = \sum_{j=1}^{N} \frac{\beta_{j}}{\lambda_{j}}\ket{u_{j}}
\end{equation}
for an input matrix $A$
\begin{equation}
    A = \sum_{j=1}^{N}\lambda_{j}\ket{u_{j}}\bra{u_{j}}
\end{equation}
and an input state $\ket{b}$
\begin{equation}
    \ket{b} = \sum_{j=1}^{N}\beta_{j}\ket{u_{j}}.
\end{equation}
Here $\lambda_{j}$ and $\ket{u_{j}}$ are eigenvalues and eigenvectors of the input matrix $A$.
The HHL algorithm produces a following quantum state
\begin{equation}
    \label{HHLoutcome}
    \ket{\Psi} = 
    \sum_{j=1}^{N}\beta_{j}\ket{u_{j}}\biggl( \sqrt{1 - \frac{C^{2}}{\lambda_{j}^{2} }}\ket{0}_{a} + \frac{C}{\lambda_{j}}\ket{1}_{a} \biggl).
\end{equation}
After measurement of ancilla qubit, this state transforms into one of two quantum states
\begin{equation}
    \label{x0}
    \ket{x_{0}} = \sum_{j=1}^{N}\beta_{j}\sqrt{1 - \frac{C^{2}}{\lambda_{j}^{2} }}\ket{u_{j}}, 
    \quad \text{when we measure } a = 0,
\end{equation}
\begin{equation}
    \label{x1}
    \ket{x_{1}} =  \sum_{j=1}^{N}\beta_{j}\frac{C}{\lambda_{j}}\ket{u_{j}}, 
    \quad \text{when we measure } a = 1,
\end{equation}
which we provide here in unnormalized form for simplicity (we will restore state norms further). The original HHL algorithm tells, that the state (\ref{x1}) is a solution of the linear system up to a constant $C$ (compare to (\ref{quantumx})). At the same time, (\ref{x0}) is the outcome when the algorithm fails to solve the linear system problem in a sense of original HHL algorithm. In the following sections we will show that, under particular condition, the state (\ref{x0}) provides the same observable value $<M>$ as a state (\ref{x1}).

\subsection{Observable value for different ancilla outcomes}

To have a quantum speed up, the result of the HHL algorithm is used in two ways: it is used as an input to another quantum algorithm, or it is used to measure some quantity. Here we concentrate on the second way and consider an observable value, represented with a measurement operator $M$. The result of measuring this operator on a final state (\ref{x1}) is a correct outcome of the algorithm. At the same time, measurement on a state (\ref{x0}) is considered a failure. 
In this section we are going to find a connection of observable values for states (\ref{x0}) and (\ref{x1}).

Let us rewrite eq. (\ref{x0}) in a form of a linear system solution:
\begin{equation}
    \ket{x_{0}} = \Tilde{A_{C}}^{-1}\ket{b},
\end{equation}
where $\Tilde{A_{C}}^{-1}$ denotes an inverse matrix of some linear system. The form of this matrix can be derived from (\ref{x0}):
\begin{eqnarray}
    \Tilde{A_{C}}^{-1}\ket{b} 
    = \sum_{j=1}^{N}\beta_{j}\sqrt{1 - \frac{C^{2}}{\lambda_{j}^{2} }}\ket{u_{j}} 
    = \sum_{k}\sum_{j=1}^{N}\beta_{j}\sqrt{1 - \frac{C^{2}}{\lambda_{k}^{2} }}\ket{u_{k}}\bra{u_{k}}\ket{u_{j}}
    = \sum_{k}\sqrt{1 - \frac{C^{2}}{\lambda_{k}^{2} }}\ket{u_{k}}\bra{u_{k}}\sum_{j}\beta_{j}\ket{u_{j}},
\end{eqnarray}
so a matrix form of $\Tilde{A_{C}}^{-1}$ is 
\begin{equation}
    \Tilde{A_{C}}^{-1} = \sum_{k}\sqrt{1 - \frac{C^{2}}{\lambda_{k}^{2} }}\ket{u_{k}}\bra{u_{k}}.
\end{equation}
We can connect this matrix with the matrix of the initial system as following:
\begin{eqnarray}
    \label{tildeAc}
    \Tilde{A_{C}}^{-1} 
    = \sum_{k}\frac{C}{\lambda_{k}}\sqrt{ \frac{\lambda_{k}^{2}}{C^{2}} - 1 }\ket{u_{k}}\bra{u_{k}} 
    = \sum_{j}\frac{C}{\lambda_{j}}\ket{u_{j}}\bra{u_{j}}\sum_{k}\sqrt{ \frac{\lambda_{k}^{2}}{C^{2}} - 1 }\ket{u_{k}}\bra{u_{k}} 
    = A_{C}^{-1} D,
\end{eqnarray}
where $A_{C}^{-1} = \sum_{j}\frac{C}{\lambda_{j}}\ket{u_{j}}\bra{u_{j}}$ is a normalized inverse of the initial linear system matrix and 
\begin{equation}
    D = \sum_{k}\sqrt{ \frac{\lambda_{k}^{2}}{C^{2}} - 1 }\ket{u_{k}}\bra{u_{k}}
\end{equation}
is some additional transform.

Having connection (\ref{tildeAc}), we can establish connection between observable values for two ancilla qubit outcomes. If we measure an observable $M$ on a state (\ref{x0}), we obtain a value
\begin{equation}
    \label{M0}
    \bra{x_{0}} M \ket{x_{0}} = \bra{b} (A_{C}^{-1}D)^{\dagger} M A_{C}^{-1}D \ket{b} = \bra{b} D^{\dagger}A_{C}^{-1\dagger} M A_{C}^{-1}D \ket{b} = \bra{x_{1}} D^{\dagger} M D \ket{x_{1}},
\end{equation}
and we measure an observable $M$ on a state (\ref{x1}), we obtain a value
\begin{equation}
    \label{M1}
    \bra{x_{1}} M \ket{x_{1}} = \bra{b} A_{C}^{-1\dagger} M A_{C}^{-1} \ket{b}.
\end{equation}
The identity (\ref{M0}) is valid, because $[A_{C}^{-1}, D] = 0$ as both matrices are diagonal at the same basis. 

The equation (\ref{M0}) shows that the result of measurement the observable $M$ on the quantum state (\ref{x0}) is equal to measurement of another observable $D^{\dagger}MD$ on the correct answer (\ref{x1}). In the following section we are going to transform identity for (\ref{M0}) to extract (\ref{M1}) from (\ref{M0}) and to investigate on the remainder part. 

\subsection{Connection of observable values for different ancilla outcomes}

To derive a connection between (\ref{M1}) and (\ref{M0}), let us introduce a new form for a matrix $D$ as follows
\begin{equation}
    \label{DDelta}
    D = \sum_{j}\biggl( 1 - (1 - \sqrt{\frac{\lambda_{k}^{2}}{C^{2}} - 1}) \biggl)\ket{u_{j}}\bra{u_{j}} = I - \Delta,
\end{equation}
where we introduced an auxiliary matrix $\Delta = \sum_{j}(1 - \sqrt{\frac{\lambda_{k}^{2}}{C^{2}} - 1})\ket{u_{j}}\bra{u_{j}}$. 
Let us prove a couple of identities for matrices $A_{C}$, $D$ and $\Delta$.
\begin{lemma}
    \label{lemma1}
    $\Delta^{2} = A_{C}^{2} - 2D$.
\end{lemma}
\begin{proof}
    \begin{eqnarray}
        \Delta^{2} = \sum_{j}(1 - \sqrt{\frac{\lambda_{j}^{2}}{C^{2}} - 1})^{2}\ket{u_{j}}\bra{u_{j}} = \sum_{j}\biggl( 1 - 2\sqrt{\frac{\lambda_{j}^{2}}{C^{2}} - 1} + (\frac{\lambda_{j}^{2}}{C^{2}} - 1) \biggl)\ket{u_{j}}\bra{u_{j}}
        \\
        = \sum_{j}\frac{\lambda_{j}^{2}}{C^{2}}\ket{u_{j}}\bra{u_{j}} - 2\sum_{j}\sqrt{ \frac{\lambda_{j}^{2}}{C^{2}} - 1 } \ket{u_{j}}\bra{u_{j}} = A_{C}^{2} - 2D.
    \end{eqnarray}
\end{proof}
\begin{lemma}
    \label{lemma2}
    $D^{2} = A_{C}^{2} - I$ .
\end{lemma}
\begin{proof}
    \begin{equation}
        D^{2} = \sum_{k}\biggl(\frac{\lambda_{k}^{2}}{C^{2}} - 1 \biggl)\ket{u_{k}}\bra{u_{k}} = \sum_{k}\frac{\lambda_{k}^{2}}{C^{2}}\ket{u_{k}}\bra{u_{k}} -  \sum_{k}\ket{u_{k}}\bra{u_{k}} = A_{C}^{2} - I.
    \end{equation}
\end{proof}

Using the $\Delta$-matrix form transforms (\ref{M0}) into a sum of two parts
\begin{equation}
    \label{x1DMDx1}
    \bra{x_{1}} D^{\dagger} M D \ket{x_{1}} 
    = \bra{x_{1}} M \ket{x_{1}} - \bra{x_{1}} M\Delta \ket{x_{1}} - \bra{x_{1}} \Delta M \ket{x_{1}} + \bra{x_{1}} \Delta M \Delta \ket{x_{1}} 
    = \bra{x_{1}} M \ket{x_{1}} +  \bra{x_{1}} \delta M \ket{x_{1}},
\end{equation}
where we separated an error operator 
\begin{equation}
    \label{errorM}
    \delta M = \Delta M \Delta - [M, \Delta]_{+}, \quad [M, \Delta]_{+} = M\Delta + \Delta M.
\end{equation}
Here we can see a part of (\ref{M1}) in (\ref{M0}), but the error term (\ref{errorM}) contains sum of terms with auxiliary matrix $\Delta$. We are going to simplify this through several transforms and backward application of (\ref{DDelta}) identity. 
For convenience, let us denote a commutator 
\begin{equation}
    \label{R}
    [M,\Delta] = M\Delta - \Delta M = R.
\end{equation}
Using this commutator, we can rewrite the error term (\ref{errorM}) in two forms, and then sum up results. These two forms are following:
\begin{enumerate}
    \item First form:
        \begin{equation}
            \Delta M \Delta = (M\Delta - R)\Delta = M\Delta^{2} - R\Delta,
        \end{equation}
        \begin{equation}
            [M, \Delta]_{+} = M\Delta + \Delta M = 2M\Delta - R.
        \end{equation}
        Using definition (\ref{DDelta}) and lemma (\ref{lemma1}), we obtain
        \begin{eqnarray}
            \label{deltaM1}
            \notag
            \delta M = M\Delta^{2} - R\Delta - 2M\Delta + R = M \Delta^{2} - 2M\Delta + R(I - \Delta) \\
            = M(A_{C}^{2} - 2D) - 2M(I - D) + RD = MA_{C}^{2} - 2M + RD.
        \end{eqnarray}
    \item Second form:
        \begin{equation}
            \Delta M \Delta = \Delta(M\Delta + R) = \Delta^{2} M + \Delta R,
        \end{equation}
        \begin{equation}
            [M, \Delta]_{+} = M\Delta + \Delta M = 2\Delta M + R.
        \end{equation}
        Using definition (\ref{DDelta}) and lemma (\ref{lemma1}), we obtain
        \begin{eqnarray}
            \label{deltaM2}
            \notag
            \delta M = \Delta^{2}M + \Delta R - 2\Delta M - R = \Delta^{2} M - 2\Delta M - (I-\Delta)R \\
            = (A_{C}^{2} - 2D)M - 2(I - D)M - RD = A_{C}^{2}M - 2M - DR.
        \end{eqnarray}
\end{enumerate}
Combining (\ref{deltaM1}) and (\ref{deltaM2}) lead to the following identity
\begin{equation}
    \label{deltaM}
    \delta M = \frac{1}{2}(\delta M + \delta M) = MA_{C}^{2} + A_{C}^{2}M - 2M + \frac{1}{2}[R,D]
\end{equation}
\begin{lemma}
    \label{lemma3}
    $MA_{C}^{2} + A_{C}^{2}M = [[M, A_{C}]A_{C}]$
\end{lemma}
\begin{proof}
\begin{equation}
    \label{MA2}
    MA_{C}^{2} = MA_{C}^{2} - A_{C}MA_{C} + A_{C}MA_{C} = (MA_{C} - A_{C}M)A_{C} + A_{C}MA_{C} = A_{C}MA_{C} + [M,A_{C}]A_{C}
\end{equation}
\begin{equation}
    \label{A2M}
    A_{C}^{2}M = A_{C}^{2}M + A_{C}MA_{C} - A_{C}MA_{C} = A_{C}(A_{C}M - MA_{C}) + A_{C}MA_{C} = A_{C}MA_{C} - A_{C}[M,A_{C}]
\end{equation}
\begin{equation}
    \label{K1}
    MA_{C}^{2} + A_{C}^{2}M = \biggl[ \text{using (\ref{MA2}) and (\ref{A2M})} \biggl] = 2 A_{C}MA_{C} + [M,A_{C}]A_{C} - A_{C}[M,A_{C}] = 2 A_{C}MA_{C} + [[M, A_{C}]A_{C}]
\end{equation}
\end{proof}
\noindent Using lemma (\ref{lemma3}) and (\ref{deltaM}), we obtain
\begin{equation}
    \label{deltaMprefinal}
    \delta M = A_{C} M A_{C} - 2M + \frac{1}{2}\biggl( [M, A_{C}]A_{C} - A_{C}[M, A_{C}] \biggl) = A_{C} M A_{C} - 2M + \frac{1}{2}[[M, A_{C}]A_{C}] + \frac{1}{2}[R,D]
\end{equation}
We can get rid of the $R$ matrix in the (\ref{deltaMprefinal}) using the following identity
\begin{lemma}
    \label{lemma4}
    $[R, D] = -[[M, D],D]$ 
\end{lemma}
\begin{proof}
We use definitions (\ref{DDelta}) and (\ref{R}):
\begin{equation}
    [R,D] = [[M,\Delta],D] = [[M,I - D],D] = -[[M,D],D]
\end{equation}
\end{proof}
Finally, using lemma (\ref{lemma4}) and lemma (\ref{lemma2}), we obtain
\begin{equation}
    \label{deltaMfinal}
    \delta M = A_{C} M A_{C} - 2M + \frac{1}{2}[[M, A_{C}]A_{C}] - \frac{1}{2}[[M, \sqrt{A_{C}^{2} - I}],\sqrt{A_{C}^{2} - I}]
\end{equation}
Now we have everything to formulate a connection between $\bra{x_{0}}M\ket{x_{0}}$ and $\bra{x_{1}}M\ket{x_{1}}$ explicitly. We use identities (\ref{M0}), (\ref{x1DMDx1}) and (\ref{deltaMfinal}) and $A_{C}\ket{x_{1}} = \ket{b}$ to formulate a theorem:
\begin{theorem}
    \label{maintheorem}
    For the HHL algorithm with an input matrix $A$ and an input vector $\ket{b}$, results of measuring the observable $M$ on the output state for ancilla qubit outcomes equal 0 and 1 are connected in the following way
    \begin{equation}
        \label{Mx0}
        \bra{x_{0}} M \ket{x_{0}} = \bra{b} M \ket{b} - \bra{x_{1}} M \ket{x_{1}} 
        + \bra{x_{1}}K\ket{x_{1}}
    \end{equation}
    where
    \begin{equation}
        \label{K}
        K = K_{1} - K_{2}
    \end{equation}
    \begin{equation}
        \label{K1}
        K_{1} = \frac{1}{2} [[M, A_{C}]A_{C}]
    \end{equation}
    \begin{equation}
        \label{K2}
        K_{2} = \frac{1}{2} [[M, \sqrt{A_{C}^{2} - I}],\sqrt{A_{C}^{2} - I}]
    \end{equation}
\end{theorem}
\begin{corollary}
    \label{postselectionfreecondition}
    If the error commutator $K = 0$, then results of measuring the observable $M$ on the output state for ancilla qubit outcomes equal 0 and 1 are connected in the following way
    \begin{equation}
        \label{Mx0}
        \bra{x_{0}} M \ket{x_{0}} = \bra{b} M \ket{b} - \bra{x_{1}} M \ket{x_{1}}
    \end{equation}
    and the HHL algorithm is postselection-free.
\end{corollary}
If commutators $K_{1} = 0$ and $K_{2} = 0$, then the HHL algorithm is postselection-free - 
the input vector $\ket{b}$ is given by assumption of the HHL algorithm (\cite{Harrow2009, Aaronson2015}) and the expectation value $\bra{b} M \ket{b}$ can be obtained at will. Thus, we can obtain an expectation value $\bra{x_{1}} M \ket{x_{1}}$ even if we measured $\ket{0}$ on the ancilla and obtained an output vector $\ket{x_{0}}$. 
In the following section we will show a practically-interesting example of an input matrix, which satisfies the postselection-free condition.

Previously we worked with unnormalized state for simplicity. If we restore norms of quantum state, which occur after ancilla measurement, we obtain a following connection of expectation values 
\begin{equation}
    \label{M estimate}
    \bra{x_{1}} M \ket{x_{1}} = \frac{1}{Pr(a=1)}
    \biggl( 
    \bra{b} M \ket{b} - Pr(a=0)\bra{x_{0}} M \ket{x_{0}}
    \biggl)
\end{equation}
where $\ket{x_{0}}$ and $\ket{x_{1}}$ are now normalized quantum states. This identity follows from (\ref{Mx0}) with use of a fact, that after ancilla qubit measurement a state of the input register normalizes as $\ket{x_{i}}/\sqrt{Pr(a=i)}$ with $i = 0,1$. A more detailed analysis of normalization influence is provided in Appendix section. 

%
%

\section{Explicit example}

To demonstrate the applicability of the postselection-free idea, 
we provide a particular example of a matrix, which is important in solving many practical problems, e.g., this matrix appears in solving differential equations (e.g. see \cite{Cai2013}). The matrix has a following form
\begin{equation}
    \label{tridiag}
    A = 
    \begin{pmatrix}
        a & b & 0 & \dots & 0 & 0 & 0 \\
        b & a & b & \dots & 0 & 0 & 0 \\
        0 & b & a & \dots & 0 & 0 & 0 \\
        \vdots & \vdots & \vdots & \ddots & \vdots & \vdots & \vdots \\
        0 & 0 & 0 & \dots & a & b & 0 \\
        0 & 0 & 0 & \dots & b & a & b \\
        0 & 0 & 0 & \dots & 0 & b & a \\
    \end{pmatrix}
    =
    a\sum_{k=0}^{N-1}\ket{k}\bra{k} + b\sum_{k=0}^{N-2}(\ket{k+1}\bra{k} + \ket{k}\bra{k+1})
\end{equation}
where $a$ and $b$ are dimensionless parameters which come from a problem under the scope. In the following we will keep these parameters (as well as observable values) dimensionless for convenience, although their physical units can be assigned once the original problem - i.e., where the matrix came from - is specified.

To illustrate the postselection-free work of the HHL algorithm, we first here provide a simple explicit example for a system of 2 linear equation with particular values of $a = 1.5$ and $b = 0.5$, which we take from \cite{Cao2012}. The matrix then reads as follows
\begin{equation}
    A
    =
    \begin{pmatrix}
        \label{MatrixA2}
        1.5 & 0.5 \\ 0.5 & 1.5
    \end{pmatrix}
\end{equation}
This matrix can be rewritten in a form $A = 1.5I + 0.5X$
with 
$X
= 
\begin{pmatrix}
 0 & 1 \\ 1 & 0    
\end{pmatrix}$
We take an observable $M = X$ to meet the postselection-free conditions. Then, we have that $[M,A] = 1.5[X, I] + 0.5[X,X] = 0$ and the first part of the commutator (\ref{K}) vanishes. For a second part of the commutator, we have $\sqrt{A^{2} - I} = \frac{\sqrt{3}}{2}(I + X)$, which again commutes with $M$: $[M, \sqrt{A^{2} - I}] = \frac{\sqrt{3}}{2}[X,I] + \frac{\sqrt{3}}{2}[X,X] = 0$, and thus the second part of the commutator (\ref{K}) also vanishes. 
The postselection-free condition is met in this case. We provide further details on numerical calculation of this case in Appendix A.

Next, we prove that this matrix commutes with an observable of the form $M = X \otimes X \otimes \dots \otimes X$. 
To prove this statement, we rewrite the observable in the following form
\begin{equation}
    \label{Mx}
    M = X \otimes X \otimes \dots \otimes X = \sum_{k=0}^{N-1}\ket{k}\bra{N-1-k}
\end{equation}
The commutator then takes the form
\begin{eqnarray}
    [A,M] = a[I,M] + b\sum_{k_{1}=0}^{N-2}\sum_{k_{2}=0}^{N-1}
    \biggl(
    \ket{k_{1}}\bra{k_{1}+1}\ket{k_{2}}\bra{N-1-k_{2}} - 
    \ket{k_{2}}\bra{N-1-k_{2}}\ket{k_{1}}\bra{k_{1}+1} + \\
    \ket{k_{1}+1}\bra{k_{1}}\ket{k_{2}}\bra{N-1-k_{2}} - 
    \ket{k_{2}}\bra{N-1-k_{2}}\ket{k_{1}+1}\bra{k_{1}}
    \biggl)
\end{eqnarray}
Using that $[I,M] = 0$ and orthogonality of quantum basis states $\bra{k_{i}}\ket{k_{j}} = \delta_{ij}$, we obtain 
\begin{equation}
    [A,M] = 
    b\sum_{k=0}^{N-2}
    \biggl(
    \ket{k}\bra{N-2-k} - \ket{N-2-k}\bra{k}
    \biggl)
    + 
    b\sum_{k=0}^{N-2}
    \biggl(
    \ket{k+1}\bra{N-1-k} - \ket{N-1-k}\bra{k+1}
    \biggl)
\end{equation}
Each of the two sums is equal to zero. To show this, we rearrange the summation index in every second term of each sum as follows
\begin{eqnarray}
    \notag
    &\sum_{k=0}^{N-2}\ket{k}\bra{N-2-k} - \sum_{k=0}^{N-2}\ket{N-2-k}\bra{k} 
    = 
    \biggl[k_{new} = N-2-k \biggl] 
    = \\
    &\sum_{k=0}^{N-2}\ket{k}\bra{N-2-k} - \sum_{k_{new}=0}^{N-2}\ket{k_{new}}\bra{N-2-k_{new}} 
    = 0
\end{eqnarray}
\begin{eqnarray}
    \notag
    &\sum_{k=0}^{N-2}\ket{k+1}\bra{N-1-k} - \sum_{k=0}^{N-2}\ket{N-1-k}\bra{k+1}
    = 
    \biggl[k_{new} +1 = N-1-k \biggl] 
    = \\
    &\sum_{k=0}^{N-2}\ket{k+1}\bra{N-1-k} - \sum_{k_{new}=0}^{N-2}\ket{k_{new}+1}\bra{N-1-k_{new}} 
    = 0
\end{eqnarray}
which proves that $[M,A] = 0$ for the matrix (\ref{tridiag}) and the observable (\ref{Mx}). 

\begin{figure}[ht]
	\centering
	\includegraphics[width=0.99\linewidth]{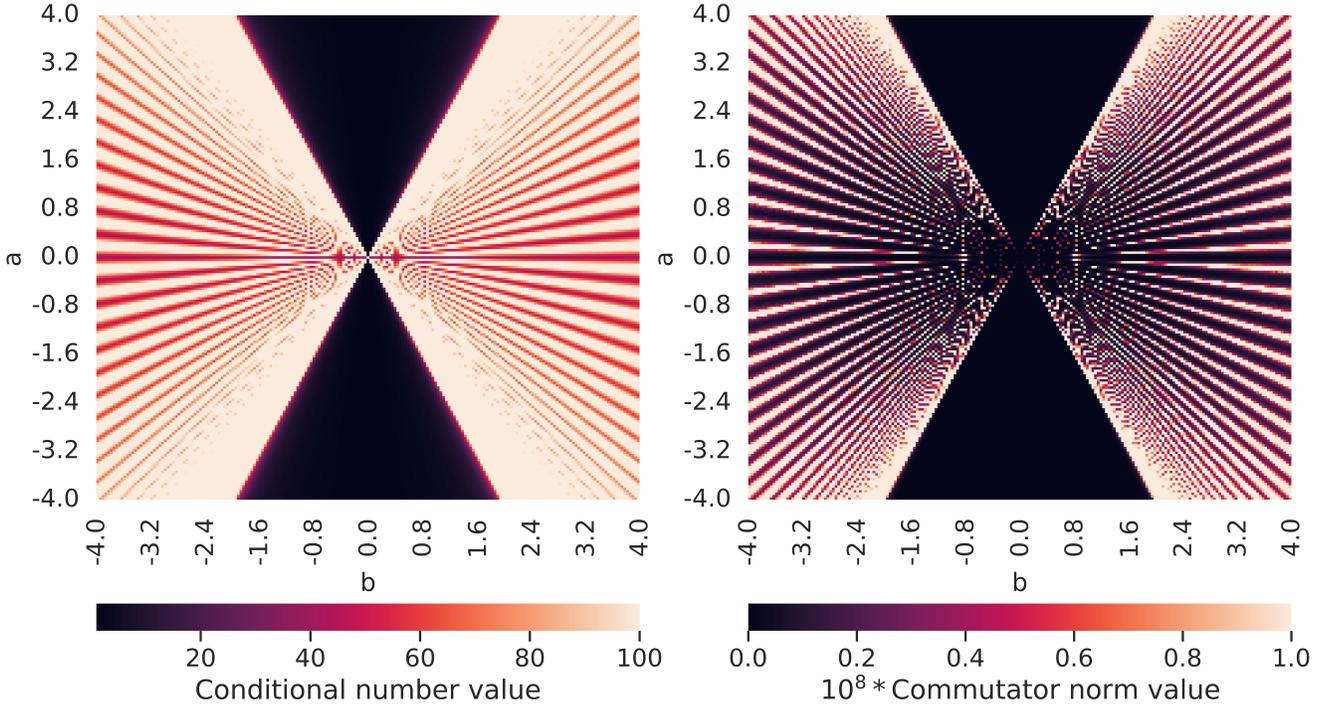}
	\caption{Heatmaps of conditional values of an input matrix $A$ (left) and a commutator (\ref{K}) values (right) over dimensionless parameters $a$ and $b$ from the input matrix. A dimension of the input matrix is $2^{6}$. Upper bound for conditional numbers is chosen $100$ and commutator Frobenius norm values are multiplied by $10^{8}$ for clarity.}
	\label{fig:kappa and K plot}
\end{figure}

The second summand of the commutator (\ref{K}) is more complicated to solve exactly, so here we resort to numerical analysis. To check the result, we calculated conditional number values of a tridiagonal matrix and a commutator (\ref{K}) Frobenius norm for different $a$ and $b$ parameters values, where the Frobenius norm is
\begin{equation}
    ||A||_{F} = \sqrt{\sum_{i}\sum_{j}|a_{i,j}|^{2}},
\end{equation}
where $a_{i,j}$ is an $(i,j)$ element of a matrix $A$. We provide corresponding heatmaps in Fig.~\ref{fig:kappa and K plot}. We calculated an input matrix of dimension $2^{N}$ with $N = 6$ for a grid of $201$ values for parameters $a$ and $b$.

We can see that the commutator Frobenius norm has close to zero values in a region of parameters $(a,b)$. This region corresponds to close to $1$ values of the input matrix conditional numbers. 
A conditional number describes a possibility to calculate an inverse of the input matrix. The more the conditional number is, the harder to calculate the inverse of the matrix with good precision. 
In the case of tridiagonal matrix (\ref{tridiag}), as long as matrix has low conditional number, it satisfies the postselection-free condition. There is a region of parameters $a$ and $b$ (a cone in the middle of heatmaps in Fig.~\ref{fig:kappa and K plot}), where this condition is satisfied.
Outside of this parameter region, the HHL algorithm does not work well as the input matrix is not invertible, and the postselection-free condition also loses meaning.

%
%

\section{Conclusion and outlook}

We demonstrated that for an input matrix $A$ and an observable $M$, which satisfy a condition (\ref{K}), the HHL can work postselection-free. We derived outcome states of the HHL for ancilla measurement in $0$ and $1$ and calculated their expectation values $M$. We showed that expectation values deviate only by an easily-accessible constant value when the postselection-free condition is satisfied. Thus, we can extract a correct function of the linear system solution from both algorithm outcomes. We provided a practical example of an input matrix $A$ and an observable $M$, which allows running the HHL without postselection.

Our result can improve algorithms that use the HHL algorithm as a subroutine. 
The HHL algorithm is efficient when an output state is used to measure an expectation value of some observable instead of measuring output vector components (which takes $O(N)$ steps). 
For this application of the HHL outcome state, we demonstrated that it is possible to get the correct output without postselection of the ancillary qubit. 
The other way to use HHL efficiently is to use the output state as an input to another quantum algorithm. 
For example, the HHL algorithm allows solving differential equations \cite{Cao2013, Berry2014, Childs2021highprecision} and numerous machine learning problems \cite{Duan2020}. It is an open question if the postselection-free condition translates to algorithms based on the HHL.
An explicit demonstration of such translation is a subject of future research. 

Another question is finding more problems, which satisfy the postselection-free condition. The commutator relation (\ref{K}) provides a recipe for construction an input matrix $A$, given a measurement matrix $M$ and vice versa. For instance, with a fixed measurement matrix $M$, one can look for an input matrix $A$, which solves a particular problem and can be effectively simulated (in the sense of Hamiltonian simulation problem \cite{Lloyd1996, Berry2015}). Contrary, with a fixed input matrix $A$, one can look for a measurement $M$, which satisfies the postselection-free condition, is efficiently realized on the quantum device, and provides a solution for the problem under the scope. 
Finally, it is reasonable to look for matrices with postselection-free running of HHL in practical matrices, such as one described in the main text. Preliminary results during the course of the main study favors that more than 3-diagonal matrices (e.g., 5- and 7-diagonal) also satisfy postselection-free condition with the observable (\ref{Mx}).
Building explicit examples of matrices, which satisfy postselection-free condition, is another subject of future research.

\section*{Acknowledgements}
The author would like to thank W.V. Pogosov and A.V. Lebedev for useful discussions of the manuscript. The author would like to thank E. D. Kelly for his remarks which helped to improve the main result.

\renewcommand{\appendixname}{APPENDIX}
\appendix


\section{Details on a 2 by 2 matrix example}

\begin{figure}[ht]
	\[
	\Qcircuit @C=1.0em @R=1.0em
	{
	    \lstick{\ket{0_{a}}}	& \qw & \qw & \qw & \qw & \qw & \qw & \qw & \gate{R_{y}(\frac{2\pi}{2^{r}})} & \qw & \gate{R_{y}(\frac{\pi}{2^{r}})} & \qw & \qw & \qw & \meter &\rstick{\ket{m}}\\
	    \lstick{\ket{0_{0,c}}}	& \qw & \gate{H} & \qw & \qw & \ctrl{2} &  \multigate{1}{QFT^{\dagger}} & \qw & \ctrl{-1} & \qw & \qw & \qw & \multigate{2}{U^{\dagger}} & \qw \\
	    \lstick{\ket{0_{1,c}} }	& \qw & \gate{H} & \ctrl{1} & \qw & \qw & \ghost{QFT^{\dagger}} & \qw & \qw & \qw & \ctrl{-2} & \qw & \ghost{U^{\dagger}} & \qw\\
	    \lstick{\ket{b}} & \qw & \qw & \gate{exp{(iA\frac{t_{0}}{4})}} & \qw & \gate{exp{(iA\frac{t_{0}}{2})}} & \qw & \qw & \qw & \qw & \qw & \qw & \ghost{U^{\dagger}} & \qw 
	}
	\]
	\caption{A quantum circuit implementing the HHL algorithm for a system of 2 linear equations for matrix (\ref{MatrixA2})}
    \label{fig:HHLcircuit}
\end{figure}
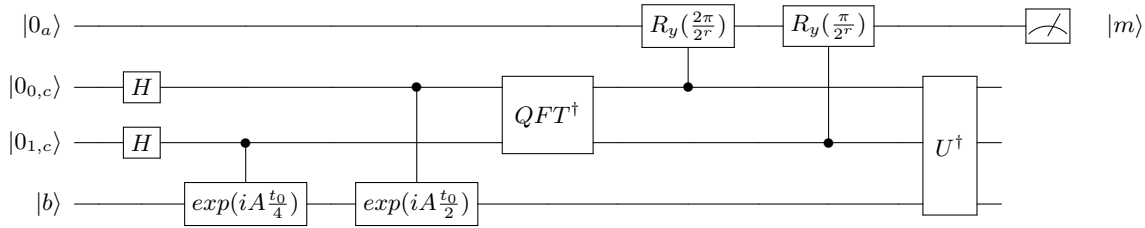

Here we provide further detail on the explicit example, described in the main text. To demonstrate the idea of the postselection-free HHL, we explicitly calculate a simple realization of the HHL algorithm, introduced in \cite{Cao2012}. Here, the HHL algorithm is used to solve a system of 2 linear equations, represented by a matrix 
\begin{equation}
    A
    =
    \begin{pmatrix}
        \label{MatrixA2}
        1.5 & 0.5 \\ 0.5 & 1.5
    \end{pmatrix}
\end{equation}
and we take an observable $M = X$.
In Fig.~\ref{fig:HHLcircuit},  we provide a quantum circuit of the HHL algorithm for this input matrix. Here $H$ is a Hadamard gate, $QFT$ is a quantum Fourier transform gate, $exp(iA t_{0})$ is a unitary evolution for matrix $A$, $R_{y}$ is a $Y$-rotation of a qubit state and $U^{\dagger}$ denotes an uncomputing gate for a sequence of gates before controlled ancilla rotations. 
The input matrix can be decomposed into a form $A = \frac{3}{2}I + \frac{1}{2}X$ and commutes with the observable $[M, A] = \frac{3}{2}[X, I] + \frac{1}{2}[X, X] = 0$. The second part of the commutator (\ref{K}) requires calculating a square root of the squared and shifted matrix $\sqrt{A^{2} - I}$. The math is the following:
\begin{eqnarray}
    A^{2} = \frac{1}{4}
    \begin{pmatrix}
    3 & 1 \\ 1 & 3    
    \end{pmatrix}
    \begin{pmatrix}
    3 & 1 \\ 1 & 3    
    \end{pmatrix}
    =
    \frac{1}{2}
    \begin{pmatrix}
    5 & 3 \\ 3 & 5    
    \end{pmatrix}
\end{eqnarray}
\begin{eqnarray}
    A^{2} - I = 
    \frac{3}{2}
    \begin{pmatrix}
    1 & 1 \\ 1 & 1    
    \end{pmatrix}   
\end{eqnarray}
\begin{eqnarray}
    \sqrt{
    \begin{pmatrix}
    1 & 1 \\ 1 & 1    
    \end{pmatrix}
    }
    =
    \frac{1}{\sqrt{2}}
    \begin{pmatrix}
    1 & 1 \\ 1 & 1    
    \end{pmatrix}
    =
    \frac{1}{\sqrt{2}}
    (I + X)
\end{eqnarray}
Finally,
\begin{eqnarray}
    \sqrt{A^{2} - I} = 
    \frac{\sqrt{3}}{2}
    (I + X)  
\end{eqnarray}
It is easy to see, that this matrix also commutes with the matrix of a chosen quantum observable: $[M, \sqrt{A^{2} - I}] = \frac{\sqrt{3}}{2}[X, I] + \frac{\sqrt{3}}{2}[X, X] = 0$.

\subsection{State vector simulator}
\begin{figure}[ht]
	\centering
	\includegraphics[width=0.99\linewidth]{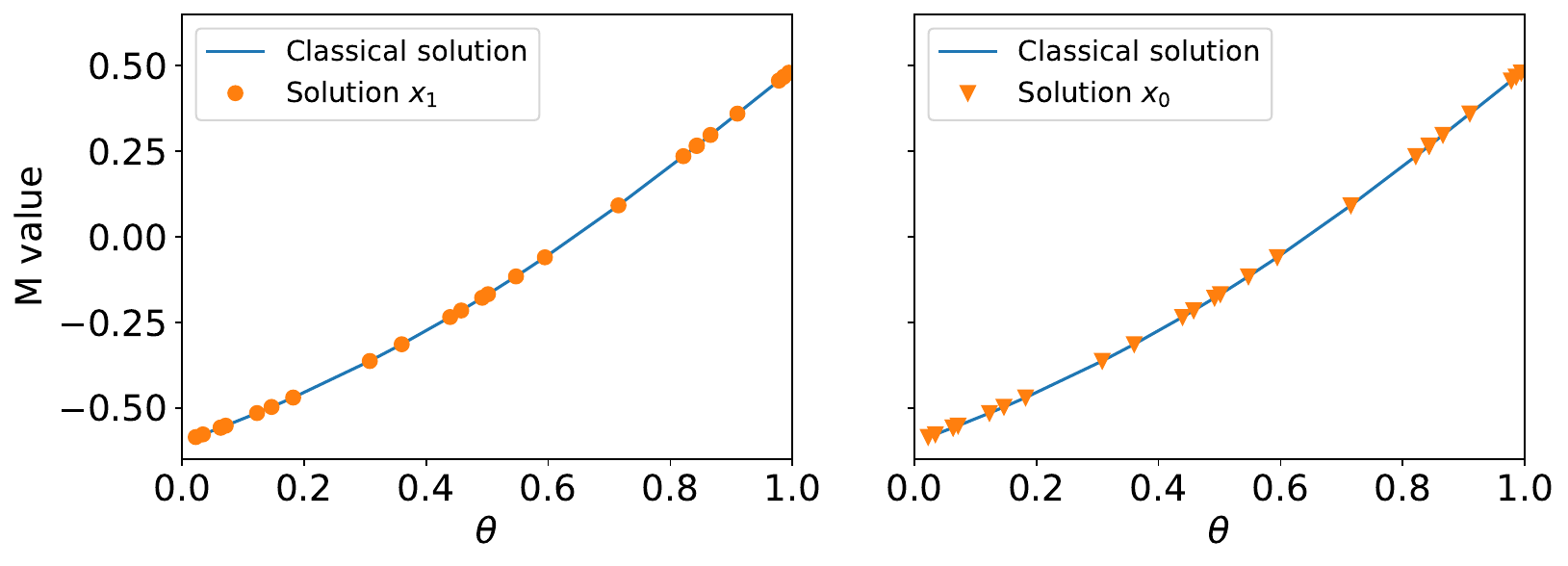}
	\caption{Expectation values of $M$ on the HHL algorithm outcomes $\ket{x_{1}}$ (left plot), $\ket{x_{0}}$ (right plot), compared to $\Vec{x}^{T} M \Vec{x}$ value (blue solid curve), where $\Vec{x}$ is a classical vector of the linear system solution. Classical solution values for different values of a parameter $\theta$ are connected with a curve to guide an eye. Parameter $\theta$ is a dimensionless parameter of the initial quantum state.}
	\label{fig:M values plot}
\end{figure}

\begin{figure}[ht]
	\centering
	\includegraphics[width=0.99\linewidth]{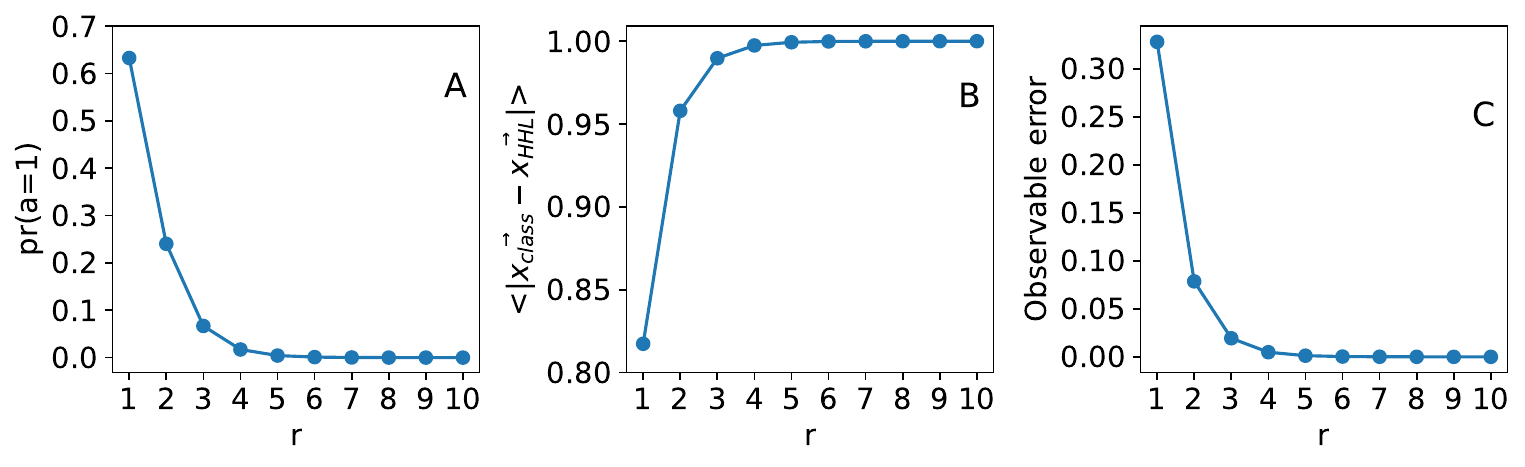}
	\caption{Dependencies of a probability to measure the ancilla in a state $\ket{1}$ (A), an average absolute difference between classical and quantum solutions (B), and a difference between observable values on classical and quantum solutions (C) on a dimensionless parameter $r$, which governs controlled rotation of the ancillary qubit in the HHL algorithm.}
	\label{fig: fidelity and probabilities}
\end{figure}

We run the algorithm for 25 random initial vectors $\ket{b} = \cos{\frac{\theta}{2}}\ket{0} + \sin{\frac{\theta}{2}}\ket{1}$ and obtained solution vectors $\ket{x_{0}}$ for ancilla measurement in state $\ket{0}$ and $\ket{x_{1}}$ for ancilla measurement in state $\ket{1}$. 
In Fig.~\ref{fig:M values plot} we provide expectation values $M_{1} = \bra{x_{1}} M \ket{x_{1}}$ and $\frac{1}{Pr(a=1)}(M_{b} - Pr(a=0)M_{0})$, where $M_{0} = \bra{x_{0}} M \ket{x_{0}}$ and $M_{b} = \bra{b} M \ket{b}$.
We compared results with a classical solution for the system of linear equations $\Vec{x}$, in particular, with a value $\Vec{x}^{T} M \Vec{x}$. From Fig.~\ref{fig:M values plot}, we see perfect coincidence with classical results and results obtained from the HHL algorithm outcomes for ancilla values $0$ and $1$. In Fig.~\ref{fig: fidelity and probabilities} we provide dependencies of probability to measure the ancilla in a state $1$, fidelity of the $\ket{x_{1}}$ with respect to classical solution and an error on observable $M$ between classical and quantum solutions on a dimensionless parameter $r$, which governs a rotation constant $C = \frac{2\pi}{2^{r}}$ in the HHL algorithm (refer to (\ref{HHLoutcome})). We can see that increasing $r$ (lowering $C$) leads to better precision of the HHL performance while decreasing the probability of measuring the ancilla qubit in a state $\ket{1}$. A low probability of ancilla measurement (thus, a low probability of the HHL correct outcome) is usually compensated with an amplitude amplification \cite{Brassard2002}. Here we demonstrate that using both output states for two ancilla measurement outcomes can give a correct expectation value of $M$ with only slight postprocessing (subtracting a constant value). This extraction of correct expectation values is possible if a postselection-free condition 
(\ref{postselectionfreecondition}) is satisfied.

\subsection{QASM simulator}

Measurement of an observable value in quantum computing requires gathering statistics on a computational basis. Thus, every observable value has a statistical error, which depends on the size of the sample. Here we analyze how statistical error influences the observable value of $M$, obtained from two cases of ancilla qubit measurement.

\begin{figure}[ht]
	\centering
	\includegraphics[width=0.75\linewidth]{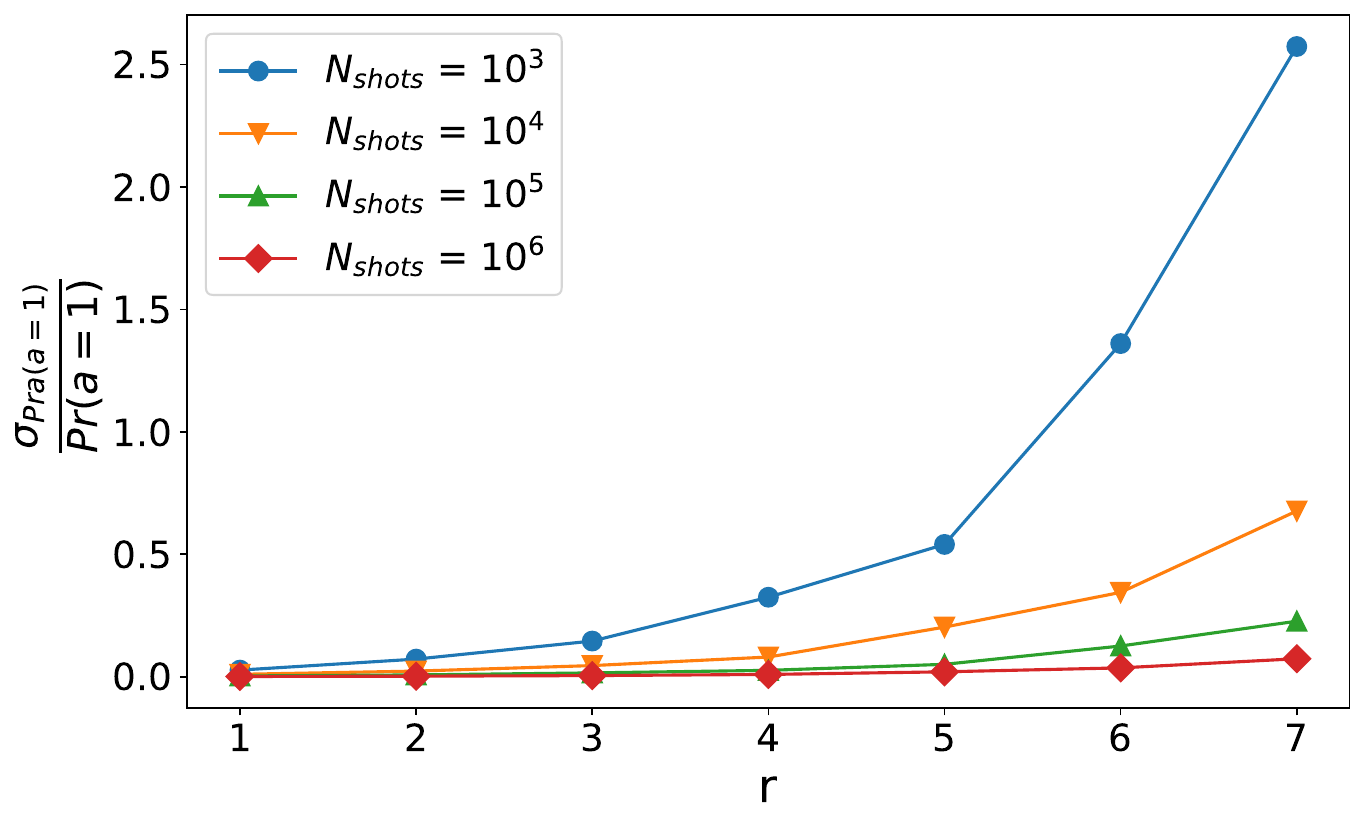}
	\caption{Dependence (on parameter r) of a standard deviation to value relation of an estimated probability to measure ancilla in a state $\ket{1}$.}
	\label{fig: variance p1 values plot}
\end{figure}

In the HHL algorithm, an important value for the algorithm result is the probability to measure ancilla in a state $\ket{1}$. This value is needed to introduce a proper normalization of $<M>$ and make it equal to a classical value of $\Vec{x}^{T} M \Vec{x}$ - for both cases of ancilla measurement we need to divide the result on an estimate of $Pr(a=1)$. In the previous section, we have seen that $Pr(a=1)$ decreases as a constant $C$ decreases while the algorithm precision increases. The smaller the $Pr(a=1)$ value, the harder it is to estimate this value with gathered statistics. For example, if $Pr(a=1)$ is of order $10^{-3}$, it is required to make about $10^{3}$ algorithm runs in average to obtain one measurement of ancilla in state $\ket{1}$. In Fig.~\ref{fig: variance p1 values plot} we provide a variance-value ration of $Pr(a=1)$ for different number of algorithm runs (shots). We can see that we need to make more runs of the algorithm to estimate $Pr(a=1)$ as algorithm precision increases (in a sense of result fidelity, provided in Fig.~\ref{fig: fidelity and probabilities} (B)). That means that a number of shots $N_{shots}$ and a constant value $C$ have a trade-off, and we need to choose them concerning requirements of algorithm results fidelity and time-consumption (if any).

\begin{figure}[ht]
	\centering
	\includegraphics[width=0.99\linewidth]{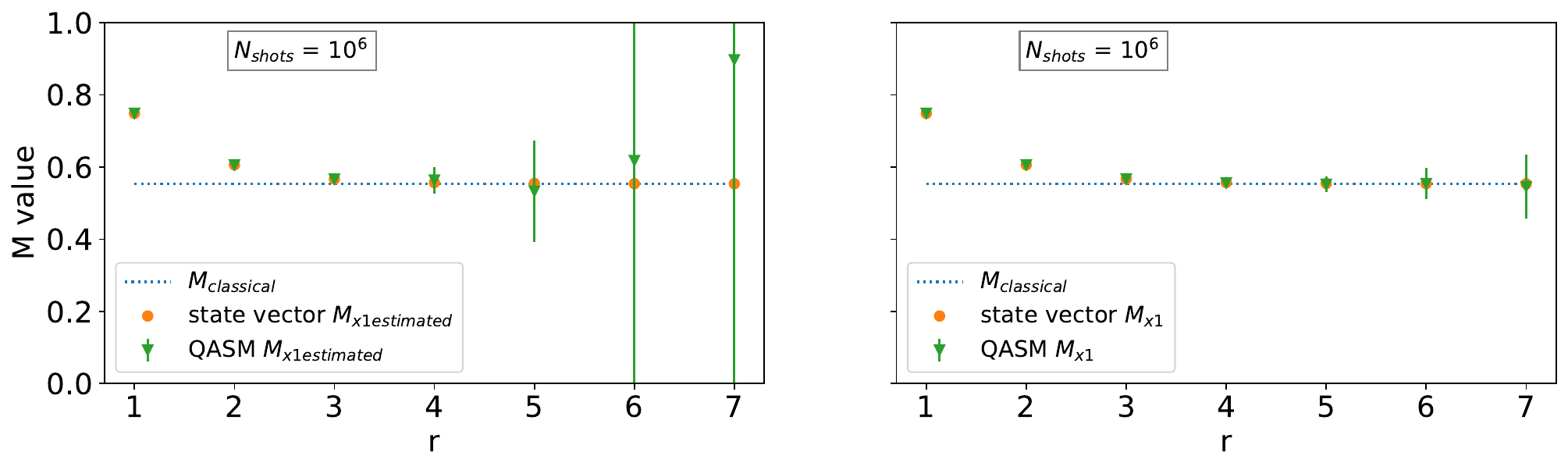}
	\caption{Dependencies (on parameter r) of estimated $M$ value on resulting HHL vectors with ancilla measured in state $\ket{0}$ (left), and ancilla measured in state $\ket{1}$. Every point is an average of 100 values, each of which is calculated by gathering statistics of $10^{6}$ shots.}
	\label{fig: estimated M values plot}
\end{figure}

For a fixed number of algorithm runs $N_{shots}$, a plot of resulting observable $M$ values for two cases of ancilla measurement is provided in Fig.~\ref{fig: estimated M values plot}. Here are provided estimates of observable values with standard deviations for different values of algorithm constant $C$ (remember that $C$ is parametrized with a parameter $r$ in the considered toy problem). 
First, we can see that for $r \geq 3$ both estimates (dots on plots) converge to a classical value $\Vec{x}^{T} M \Vec{x}$. That means that, for a particular problem, there is a sufficient value of $C$, which gives adequate precision to the answer, and taking $C$ lower does not increase the estimate precision significantly.
Second, we can see that the standard deviation (error bars on plots) is different for two ancilla measurement cases. If we measure ancilla qubit in a state $\ket{0}$ and construct an estimate of observable $M$ with equation (\ref{M estimate}), we obtain a correct estimate with a standard deviation larger than in the case of the straightforward HHL algorithm use, when ancilla is measured in state $\ket{1}$. The estimate (\ref{M estimate}) uses estimated values of $\bra{b}M\ket{b}$, $\bra{x_{0}}M\ket{x_{0}}$ and $Pr(a=1)$, each of which has statistical error. In the numerator of (\ref{M estimate}), we have two estimated values with non-zero variance, hence we have two sources of uncertainty instead of one in the case when ancilla is measured in state $\ket{1}$. As a result, the method to estimate observable $M$ value for the state $\ket{x_{0}}$, which we propose in this paper, provides a correct estimate with a price of a higher statistical error. Nonetheless, this increased statistical error is not dramatic, and, with a proper choice of a constant $C$ and a number of algorithm runs $N_{shots}$, we can obtain estimates with comparable precision.

\begin{figure}[ht]
	\centering
	\includegraphics[width=0.75\linewidth]{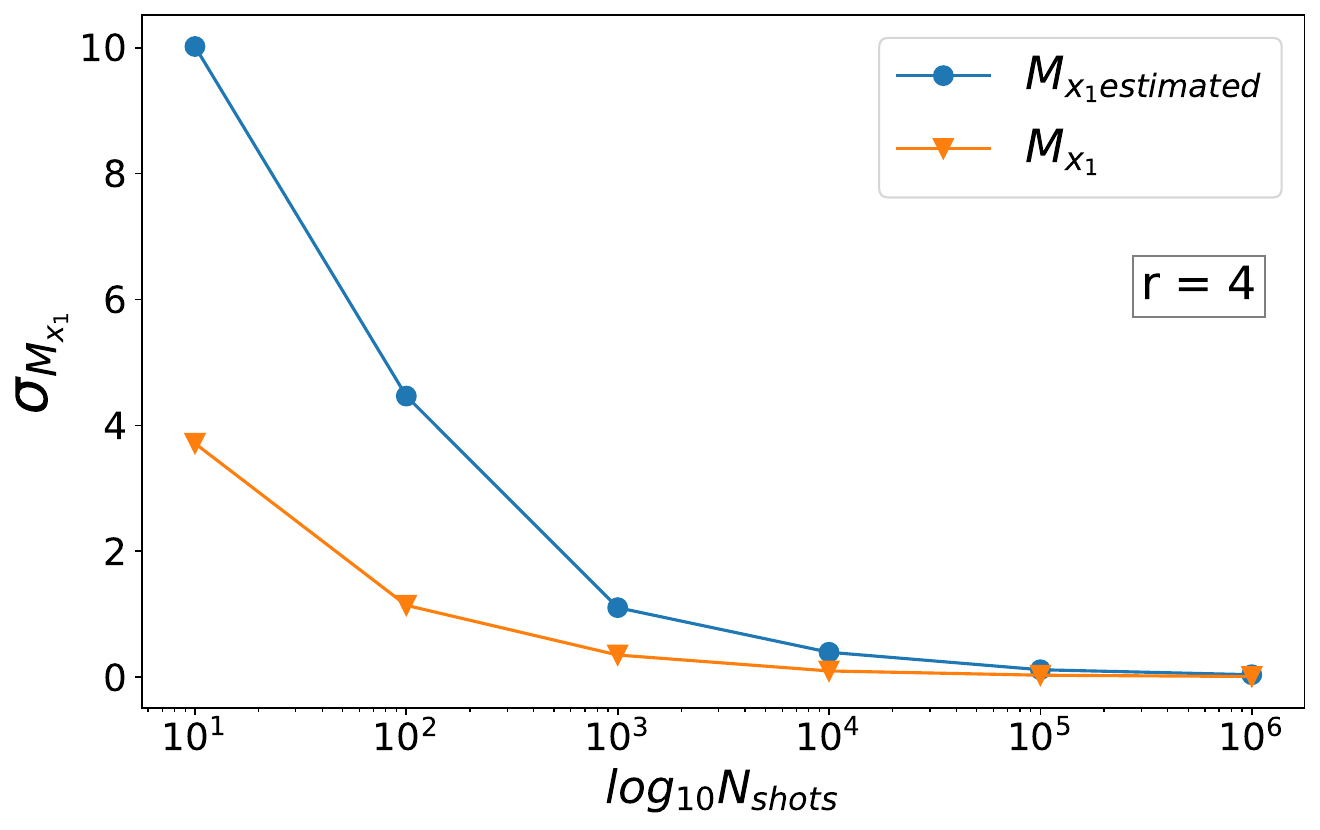}
	\caption{Dependence (on a logarithm number of shots) of standard deviations of observable value estimates for ancilla measured in a state $\ket{0}$ ($M_{x_{1} estimated}$), and for ancilla measured in a state $\ket{1}$ $(M_{x_{1}})$. Both dependencies are provided for a parameter value $r = 4$. }
	\label{fig: variance M values plot}
\end{figure}

For a fixed value of parameter $r = 4$, a plot of the standard error of observable value estimates is provided in Fig. \ref{fig: variance M values plot}. For a number of algorithm runs more than $10^{4}$, standard errors of two estimates (for ancilla measured in state $\ket{1}$ and $\ket{0}$) are of comparable value.  

\bibliographystyle{apsrev4-1}
\bibliography{ref}

\end{document}